\documentclass[twocolumn,showpacs,preprintnumbers,amsmath,amssymb]{revtex4}

\usepackage{graphicx}
\usepackage{dcolumn}
\usepackage{bm}
\begin{document}

\preprint{APS/123-QED}

\title{Large magnetoresistance at room-temperature in semiconducting polymer sandwich devices}

\author{T. L. Francis$^1$,\"{O}. Mermer$^2$,G. Veeraraghavan$^1$,M. Wohlgenannt$^2$}
\email{markus-wohlgenannt@uiowa.edu}
\affiliation{$^1$Department of Electrical and Computer Engineering, The University of Iowa, Iowa City, IA 52242-1595}%
\affiliation{$^2$Department of Physics and Astronomy, The University of Iowa, Iowa City, IA 52242-1479}%

\date{\today}

\begin{abstract}
We report on the discovery of a large, room temperature magnetoresistance (MR) effect in polyfluorene sandwich devices in weak magnetic fields. We characterize this effect and discuss its dependence on voltage, temperature, film thickness, electrode materials, and (unintentional) impurity concentration. We usually observed negative MR, but positive MR can also be achieved under high applied electric fields. The MR effect reaches up to 10\% at fields of 10mT at room temperature. The effect shows only a weak temperature dependence and is independent of the sign and direction of the magnetic field. We find that the effect is related to the hole current in the devices.

\end{abstract}

\maketitle

Organic conjugated materials have been used to manufacture promising devices such as organic light-emitting diodes (OLEDs) \cite{ElectroluminescenceReview}, photovoltaic cells \cite{PVCellReview} and field-effect transistors \cite{reviewFET}. Recently there has been growing interest in spin \cite{nature,TalianiPaper,ValyNature} and magnetic field effects \cite{Bussmann,Preprint,Bussmann2} in these materials in order to assess the possibility of using them in magnetoresistive device applications. During the study of polyfluorene (PFO) sandwich devices \cite{Preprint} we surprisingly discovered a large and intriguing magnetoresistance (MR) effect. In our best devices this MR effect reaches up to 10\% (defined as $\Delta R/R \equiv (R(B)-R(0)/R(0)$) at fields, $B=10 mT$ at room temperature. To the best of our knowledge, this constitutes a record value in bulk materials. In the following we experimentally characterize the effect. At the end of the paper, we will discuss possible mechanisms that cause the MR effect.

Our thin film sandwich devices consist of the polymer PFO (poly(9,9-dioctylfluorenyl-2,7-diyl) end capped with N,N-Bis(4-methylphenyl)-4-aniline, see Fig. 1 inset) sandwiched between a top and bottom electrode. The polymer was purchased from American Dye Source, Inc (ADS), as well as from H. W. Sands (HWS) and was used as received. The polymeric film was fabricated by spin-coating from toluene solution at 2000 rpm. For varying the film thickness, different concentrations were used, namely 7 to 30 mg/ml. The bottom electrode consisted of either indium-tin-oxide (ITO) covered glass, poly(3,4-ethylenedioxythiophene) poly(styrenesulfonate) (PEDOT) spin-coated on top of ITO, or Au evaporated onto a glass slide. The top contact, either Al, Ca (covered by a capping layer of Al) or Au, was evaporated through a shadow mask (active area: 1 $mm^2$) at a base pressure of $10^{-6}$ mbar. All manufacturing steps were performed inside a nitrogen glove-box. The MR measurements were performed with the sample mounted on the cold finger of a closed-cycle He cryostat located between the poles of an electromagnet. The MR was determined by measuring the current at a constant applied voltage, V.

\begin{figure}
\includegraphics[width=\columnwidth]{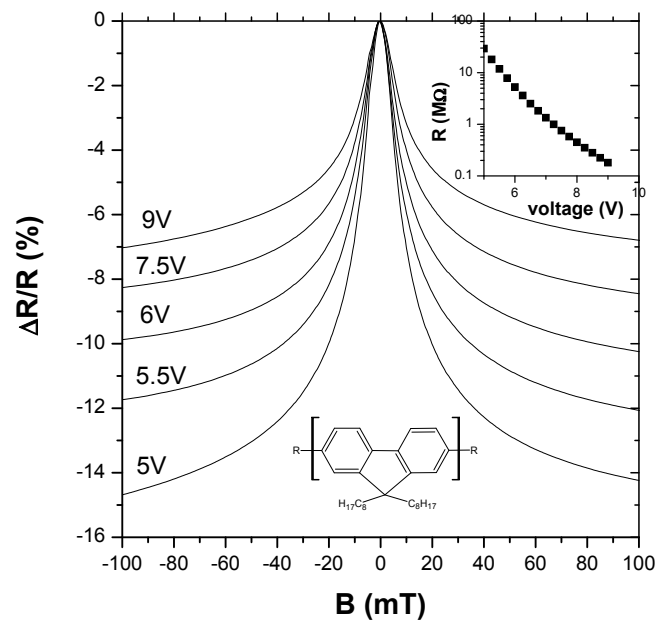}
\caption{\label{fig:Fig1} Magnetoresistance, $\Delta R/R$ curves, measured at room temperature in an ITO (30 nm)/PEDOT ($\approx$ 100nm)/PFO ($\approx$ 150 nm)/Ca ($\approx$ 50nm including capping layer) device at different voltages. The inset shows the device resistance as a function of the applied voltage.}
\end{figure}

Fig.~\ref{fig:Fig1} shows measured MR traces in a PFO sandwich device (details are given in the caption) at room-temperature at different V. We found that the measured MR traces are independent of the angle between film plane and applied magnetic field. All measurements shown were performed with an in-plane magnetic field.

\begin{figure}
\includegraphics[width=\columnwidth]{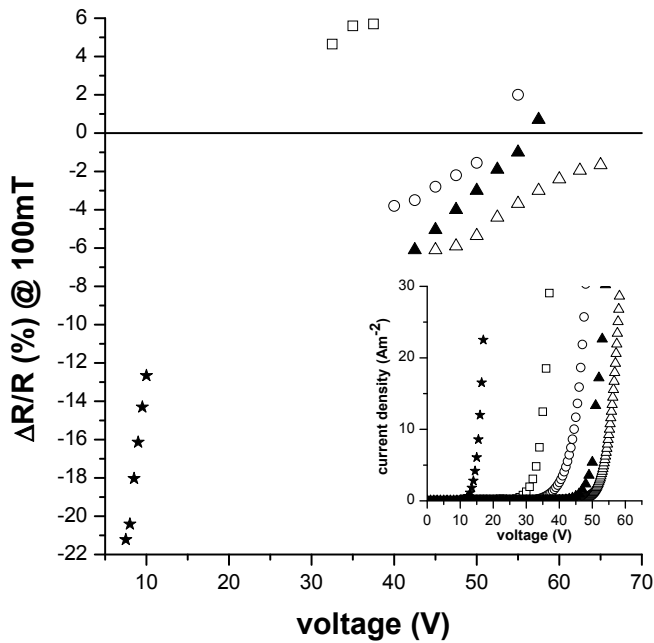}
\caption{\label{fig:Fig2} Dependence of $\Delta R/R$ at 100 mT and 200 K on the device voltage in a variety of PFO devices with different electrode materials. The inset shows the current-voltage characteristics of these devices. $\star$ is for PEDOT/PFO ($\approx$ 150 nm)/Ca, $\circ$ is for ITO/PFO ($\approx$ 140 nm)/Ca, $\blacktriangle$ is for ITO/PFO ($\approx$ 150 nm)/Au, $\vartriangle$ is for Au/PFO ($\approx$ 150 nm)/Ca, $\square$ is for ITO/PFO ($\approx$ 100 nm)/Ca at high applied voltages.}
\end{figure}

Fig.~\ref{fig:Fig2} shows the dependence of the magnitude of the MR effect at 100 mT and 200 K (to reduce thermal drift during measurements) on V in a variety of devices using different electrode materials with a polymer film thickness of $\approx$ 150nm (details are given in the caption). PEDOT and Ca are commonly used in OLEDs since they result in relatively small barriers for hole and electron injection, respectively. ITO and Au are other common contacts for hole injection because of their large work function. We used Ca, Al (data not shown) or Au as the top electrode material, resulting in efficient (Ca) or moderately efficient (Al) electron injection or hole-only devices (Au). The current-voltage (IV) characteristics of the measured devices are shown as an inset to Fig.~\ref{fig:Fig2}. It is seen that the IV curves are strongly non-linear as is usually the case in polymer sandwich devices. We found that both IV and MR curves do not critically depend on whether Au or ITO are used. However, using PEDOT as the anode results in a significant reduction in onset voltage and an increase in the observed MR effect. Both results can be understood considering the decrease in the hole-injection barrier, and therefore the reduction of the interface series resistance, which is the reason that PEDOT is the preferred anode material in OLEDs. \emph{Importantly, the observed MR effect is largely independent of the top electrode (electron injector) material} and occurs also in hole-only devices. This clearly indicates that the MR effect is due to hole transport, and not connected to electron transport or electron-hole recombination processes that also occur in OLED devices. Most of the data shown is measured in PEDOT/PFO/Ca devices since these showed the best long term stability.

\begin{figure}
\includegraphics[width=\columnwidth]{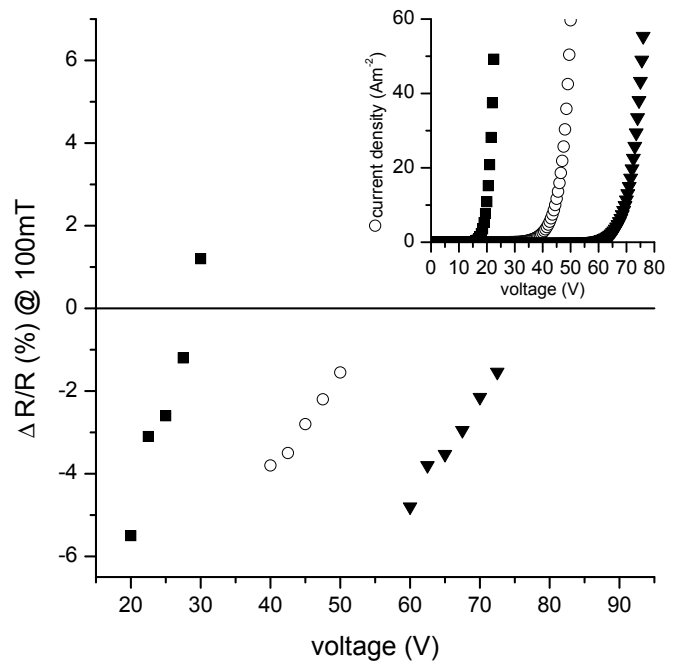}
\caption{\label{fig:Fig3} Dependence of $\Delta R/R$ at 100 mT and 200 K on the device voltage in a variety of devices with different polymer film thickness. The inset shows the IV characteristics of these devices. $\blacksquare$ is for an ITO/PFO ($\approx$ 60 nm)/Ca device, $\circ$ is for ITO/PFO ($\approx$ 140 nm)/Ca, and $\blacktriangledown$ is for ITO/PFO ($\approx$ 300 nm)/Ca.}
\end{figure}

Fig.~\ref{fig:Fig3} shows the dependence of the magnitude of the MR effect in ITO/PFO/Ca devices with different polymer film thickness (details are given in the caption) at 100 mT and 200 K on V. We found that the onset voltage in the linear-linear IV plot in these devices is determined mostly by the PFO film thickness. Since similar results (not considering the shift in operating voltage) are achieved independent of PFO film thickness, this clearly suggests that the observed MR effect is a bulk, rather than an (electrode) interface effect. This conclusion is of course further supported by the fact that the MR effect is observed for PEDOT, ITO and Au anodes. We note that the observation that the MR effect in PEDOT devices is considerably larger than for Au and ITO devices is not in contradiction with our conclusion, but can be attributed to a reduction in interface series resistance in the device when using PEDOT.

Returning to the data in Fig.~\ref{fig:Fig1} it is seen that $\Delta R/R$ typically increases in magnitude with increasing device resistance. However, we find that the resistance of our devices decreases much faster with increasing V than does the magnitude of the MR effect. We conjecture that this weak dependence of $\Delta R/R$ over several orders of magnitude in R suggests that the "intrinsic" MR is independent of V or current, and that the weak dependence is an "artifact" due to series resistances outside of the PFO film, such as hole-injection (Schottky-like) interface series resistance. Another striking result shown in Figs.~\ref{fig:Fig2} and ~\ref{fig:Fig3} is that, in addition to negative MR, positive MR can be observed \cite{Preprint} in ITO and Au anode devices at high V. We note that we have never observed positive MR in PEDOT devices and speculate that this may be related to the significantly reduced onset voltage in this devices. We note that the applied electric fields are very large in polymer sandwich devices (typically $10^{5}$ to $10^{6} V/cm$).

\begin{figure}
\includegraphics[width=\columnwidth]{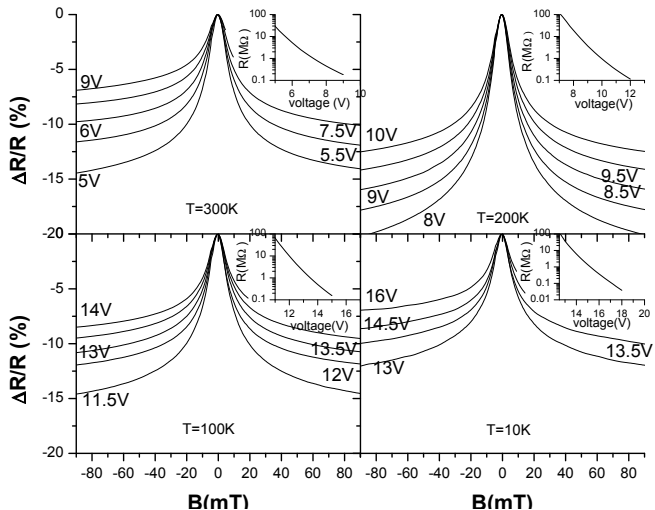}
\caption{\label{fig:Fig4} Magnetoresistance, $\Delta R/R$ curves in an PEDOT/PFO ($\approx 150 nm$)/Ca device measured at different temperatures, namely 10 K, 100 K, 200 K, and 300 K. The applied voltages are assigned. The insets show the IV characteristics at the different temperatures.}
\end{figure}

Fig.~\ref{fig:Fig4} shows MR traces in a PEDOT/PFO/Ca device for four different temperatures between 300K and 10K. We find that the magnitude and width of the MR cones are relatively insensitive to temperature. Fig.~\ref{fig:Fig4}, inset shows the IV curves at the different temperatures.

Similar experiments were also performed on devices made from other $\pi$-conjugated polymers and small molecules and will be reported elsewhere. It is important to address the question whether the MR effect may be related to (unintentional) impurities, such as left over catalysts from the polymerization reaction. Elemental analysis performed by ADS of our PFO batches showed signals for only one impurity, namely the catalyst bis(1,5-cyclooctadiene)Nickel (NiCOD) at levels less than 20 ppm. In addition we commissioned 3 new batches of PFO that have been purified to different degree after synthesis, resulting in NiCOD content of 21, 177, 683 ppm. Our MR measurements on these badges showed no significant dependence on NiCOD impurity concentration. In addition, the MR effect in devices made from HWS batches showed no significant difference compared to ADS batches. We consider this strong evidence that the MR effect is not related to (unintentional) impurities.

Finally, we want to briefly discuss possible mechanisms to explain the observed MR effect. We are familiar with the following mechanisms that cause MR: (1) Lorentz force, (2) hopping magnetoresistance \cite{EfrosBook}, (3) electron-electron interaction \cite{MRBook} and (4) weak localization \cite{WeakLocalizationMetals}. It appears that mechanisms (1) to (3) cannot explain our MR effect, because effects (1) to (3) exclusively lead to positive MR, whereas our effect is typically negative. We note that the observed MR traces closely resemble MR traces due to weak localization (negative MR) and weak antilocalization (positive MR) well known from the study of diffusive transport in metals and semiconductors \cite{WeakLocalizationMetals,WeakLocalizationFET,WLQuantumDot}. This suggests analyzing the MR data using the theory of weak localization. Such an interpretation however results in several surprising results that cast some doubt on this interpretation \cite{Preprint}. It therefore appears that a novel explanation for the observed MR effect needs to be found.

In summary, we discovered a large MR effect in PFO sandwich devices. The magnitude of the effect is several percent at fields of order 10mT and can be either positive or negative, dependent on V. The effect is independent of the sign and direction of the magnetic field, and is only weakly temperature dependent. The MR effect appears to be a bulk effect related to the hole current.

We acknowledge fruitful discussions with Profs. M. E. Flatt\'{e} and Z. V. Vardeny. This work was supported by Carver 8 50152 00 and University of Iowa, MPSFP 8 50029 55 grants and NSF ECS 04-23911.

\bibliography{magnetoresistance}

\end{document}